# The consequence of substrates of large-scale rigidity on actin network tension in adherent cells

I Manifacier[1], KM Beussman[2], SJ Han[3], NJ Sniadecki[2], I About[1], JL Milan[1]

[1] Aix Marseille Univ, CNRS, ISM, Inst Movement Sci, Marseille, France.

[2] University of Washington, Seattle, WA, USA

[3] Department of Biomedical Engineering, Michigan Technological University, Houghton, Michigan.

Email: jean-louis.MILAN@univ-amu.fr

## 1. Abstract

There is compelling evidence that substrate stiffness affects cell adhesion as well as cytoskeleton organization and contractile activity. This work was designed to study the cytoskeletal contractile activity of cells plated on microposts of different stiffness using a numerical model simulating the intracellular tension of individual cells. We allowed cells to adhere onto micropost substrates of various rigidities and used experimental traction force data to infer cell contractility using a numerical model. The model discriminates between the influence of substrate stiffness on cell tension and shows that higher substrate stiffness leads to an increase in intracellular tension. The strength of this model is its ability to calculate the mechanical state of each cell in accordance to its individual cytoskeletal structure. This is achieved by regenerating a numerical cytoskeleton based on microscope images of the actin network of each cell. The resulting numerical structure consequently represents pulling characteristics on its environment similar to those generated by the cell *in-vivo*. From actin imaging we can calculate and better understand how forces are transmitted throughout the cell.

## 2. Author Summary

Scientists at Aix-Marseille University and the University of Washington have been trying to understand the mechanical structure of human adherent cells which are found in most tissues: "We knew that adherent cells are sensible to the mechanical properties of their environment and that they generate forces accordingly. It's a very important phenomena related to development, tissue regeneration and cancer. By associating an advanced mechanical measurements technic, with microscope images of the cell structure we were able to recreate a self-optimizing computer model that indicates the amount of force transiting through the cell. Other and our results showed that cells generate higher internal tension and become larger when their surrounding becomes stiffer. Interestingly, our results also suggest that cell size is not directly associated to higher internal tension. To our surprise these results also suggests that internal tension maybe due to the superposition of at least two different phenomena. One of which appears to be influenced by adhesion conditions (how the cell is gripping to its surrounding), while the other is not. This is exciting because it changes the way we look at the cell structure and thus tackle vital subjects such as tissue regeneration and cancer."

# Acknowledgments

This work was supported from by a grant from the French Research Agency (ANR-12-BSV5-0010). The authors thank Frédéric Dubois and his research team for their help and making the LMGC90 software freely available.

# 3. Introduction

Substrate stiffness influences adherent cell behavior, both mechanically and biologically. For instance, a stiffer substrate will influence adherent cells to pull more on their surroundings (Discher 2005). Likewise variations in substrate stiffness can influence cell differentiation (Fu 2010). Several studies have investigated the biochemical aspects of mechanotransduction occurring during cell adhesion and have identified relationships between substrate stiffness, cell shape, traction forces and cell differentiation (Wang2009, Tan 2003, Fu 2010, Legant 2010, Rape 2011). Among the numerous studies on the subject of cell adhesion and mechanics, some of them proposed computational model to quantify these intracellular forces simultaneously throughout the cell.

Various 3D computational models have been developed to characterize the mechanical response of the cell under external loading or to assess the influence of the various substructures of the cytoskeleton on the overall mechanical response of the cell (McGarry2004, Barreto2012, Barreto2013, Kim2009, Mak2016, Gadilin2014, Pivkin2016, Lennon2011) and some, more precisely, to better understand how substrate stiffness influences cell contractility (Sen 2009, Dokukina 2010, Torres 2012, Borau2012, Parameswaran 2014, Fang 2016). Borau et al. 2012, showed how substrate stiffness modulates intracellular contractility using active cross-linked actin networks model (Borau 2012). In Milan et al. 2016 we investigated this issue using cells cultured on microposts of different stiffness and a 3D divided medium model (Milan 2016). The cytoskeleton was computed to act as the cells did, i.e. by exerting the same level of force as in the *in-vitro* experiments on each micropost. The main limitation of the study was that the cytoskeleton was represented by an idealized cytoskeletal structure, instead of being specific to the observed structure. Furthermore, current 3D models the observed cellular structure is either approximated to a material or ignored. The cytoskeletal structure can for instance be imaged using a confocal microscope, vertical resolution (z axis) is however too poor to allow virtual CSK reconstruction in that direction (Manifacier 2016). Other 2D discrete models have then been developed (Loosli2010, Pathak2012, Soiné2015). 2D fluorescence images where thus used to assess tension in cultured cells adhering on micropost substrates, the bias being all actin sub-structures were merged in the xy plan. In adherent cells analyzed by traction force microscopy, Soiné et al. 2016 identified predominant stress fibers and represented them as pre-stressed tensile struts directly connected to the substrate. The main limit is that the stress fibers are represented by disconnected contractile segments (direct post to post connection), which differs with the interconnected nature of the actin network.

To address this point, we developed an image-based model of adherent cells which generates a numerical cytoskeleton based on fluorescence images of the actin network (Manifacier 2016). Therefore, the resulting virtual cytoskeleton possessed the same spatial organization as its original counterpart. Furthermore, its contractility was adapted to match the level of traction forces the cell applied on its focal adhesions.

Yet a significant limit of our previous study (Manifacier 2016) was that we did not measure adhesion forces accurately. In the present work, we combine the results from in-vitro experiments on cells cultured on microposts of various stiffness levels with the image-based model to estimate intracellular forces more precisely and examine the influence of substrate stiffness on intracellular tension.

The objective of this study is to link the tensile properties of the actin network to the cell size, and substrate stiffness. For this, we used four types of micropost substrates which varied only in terms of micropost lengths. We refer to these substrates, based on micropost bending stiffness, as Very soft, Soft, Hard, and Very hard. This stiffness setup enabled us, after computation of the model, to estimate the density of tension in every part of the cell in respect to substrate stiffness. The results obtained from this study may offer a different perspective on how we can interpret actin imaging in a mechanistic way to better understand adherent cell structure.

## 4. Methods

### 4.1. In-vitro cell cultures on microposts to measure the influence of substrate stiffness on traction forces (details in supplementary data)

Human pulmonary artery endothelial cells (HPAECs, Lonza) were cultured on four different micropost substrates (Han 2012) (details in supplementary data). The deflection of a post ($\delta$) was used to determine the local traction force ($F$). All four types of substrates were made of the same PDMS with a Young's modulus of 2.5 MPa measured according to ASTM standard D412 with four different bending stiffness (11.0 nN/μm, 15.5 nN/μm, 31.0 nN/μm and 47.8 nN/μm). We categorized these stiffness values into the following designations: Very Soft (11.0 nN/μm ±2.3), Soft (15.5 nN/μm ±3.6), Hard (31.0 nN/μm ±6.2) and Very hard (47.8 nN/μm ±10) (Han 2012). The edge-to-edge spacing between microposts was therefore ~7 μm. The 7 μm spacing is assumed to be wide enough to prevent ~2 μm long contractile units from being able to perceive micropost bending stiffness (Meacci 2016). For each of the 4 adhesion conditions, 8-15 cells were imaged and analyzed for traction forces and cytoskeletal tension quantification. The cells were permeabilized using 0.5% Triton for 2 min after being allowed to spread for 14 hours on the micropost arrays. Cells were then stained with Hoechst 33342 (Invitrogen), phalloidin (Invitrogen), IgG anti-vinculin (hVin1, Sigma Aldrich), and anti-IgG antibodies (Invitrogen) with manufacturer-recommended concentration. We only selected isolated cells. These cells were therefore not able to form cadherin mediated cell-cell junctions with neighboring cells, instead they formed focal adhesions which allowed them to pull on their surrounding environment. Micropost deflection and bending stiffness were used to measure the pulling force exerted by the cell on each post (Lemmon 2005).

## 4.2. Computation of intracellular tension using an image based model of the cytoskeleton

For each observed cell, the actin images were transformed into a mechanical model of the cytoskeleton (Fig. 1) as described previously described (Manifacier 2016). The image resolution was adapted so that the cell (alone) was represented by 10 000 pixels. To build the cell model, we positioned a mechanical node, so called *actin node*, at the center of each pixel. The 10,000 actin nodes defined the mesh of the model. To numerically recompose the actin network, the actin nodes connected mechanically throughout the model with their closest neighbors. The intensity of a mechanical interaction depended locally on the actin pixel brightness. The logic is similar to Beer-Lambert Law. We made the assumption that the camera sensor measured light linearly. The light emitted within an area represented by a given pixel depends on the number of fluorescent molecules that have emitted light while the camera shutter was opened. This means that between two pixels on the same image, the difference in brightness is linearly proportional to the number of fluorescent molecules. As a consequence, we conclude that the amount of the actin network is linearly proportional to the brightness. In other words, if the brightness doubles, logically so has the local amount of actin network material and stiffness. As a result, we conclude that the brightness and stiffness are linearly proportional. This assumption enabled the model to calculate a relative local concentration of actin density. As a result, the nodes were distributed and connected based on a squared like grid. Actin nodes were then connected to each other by tensile elastic interactions (between the actin particle/nodes) which generated a pre-stressed network, to simulate the contractile nature of the actomyosin network. This contractile network then pulled on its surrounding via perfect glue interactions. These tensile interactions behaved as virtual pre-strained elastic rubber bands between all neighboring actin nodes of the model (Milan 2007, Milan 2013). Those interactions would generate a traction force that was proportional to strain or became null when the virtual rubber band slackened. As shown in the set of Eq. 1, the traction force $T(x,y)$ of interaction at (x,y) location was function to the value gap $g$ (the distance between two nodes), $g_0$ being the gap at the beginning of the simulation, the stiffness **K(x,y)** > 0 defined as a force per strain, $\boldsymbol{\varepsilon_0}$ > 0 the pre-strain in the virtual elastic rubber band, $\boldsymbol{g_v}$ is the maximal gap beyond which the interaction would not be created. Based on LMGC90 conventions *T(x,y)* is thus either null or positive when the two nodes are being pulled together to mechanically mimic the shortening of an acto-myosin bundle.

$$\begin{cases} g \in \,](1-\varepsilon_0)g_0; g_v\,] \Rightarrow T_{(x,y)} = -K_{(x,y)} (\frac{g-g_0}{g_0} + \varepsilon_0) \\ \quad g \in [0; (1-\varepsilon_0)g_0] \text{ or } g \geq g_v \Rightarrow T_{(x,y)} = 0 \end{cases} \quad (1)$$

To reproduce the contractile nature of the cell at the beginning of the simulation, we generated a positive tensile pre-strain $\boldsymbol{\varepsilon_0}$ between direct neighboring nodes. Based on experimental observation from Deguchi et al. (2006), we applied 20% of pre-strain $\boldsymbol{\varepsilon_0}$ to all actin tensile interactions to generate initial contractile inter-tensions at null deformations (Deguchi 2006). The fact that the pre-strain value was defined constant throughout the cell is a reasonable simplification since measuring pre-stress in a stress fiber cannot feasibly be done throughout the cell. On the other hand, since the model is not allowed to deform during the simulation changing the pre-strain between 10%-50% globally will not affect the results as the inverse mechanics solving method will automatically adjust for it (Eq. 5).

To limit the number of tensile interactions, the tensile interaction law was given a visibility threshold gap $\boldsymbol{g_v}$ greater than $\boldsymbol{g_0}$, the initial distance between two closest actin bodies, and lower than $\sqrt{2}\boldsymbol{g_0}$ so that diagonally disposed neighboring bodies would not be able to interact. We thus arbitrarily set $\boldsymbol{g_v}$ equal to $1.1\boldsymbol{g_0}$. In other words, these values were defined based on the geometric criteria of the mesh size alone without requiring any parametric analysis.

These tensile interactions are linked to the actin fluorescence image of the cell. In the actin image, a brighter area was considered to have a high concentration of actin and was therefore mechanically represented as being relatively stiffer than a darker area. In other words, in the actin image, dark and bright areas respectively represent low and high tension. We assume that the force intensity is proportional to visual actin density (Manifacier et al. 2016, Livne and Geiger 2016). Tensile interactions were thus given a stiffness **K(x,y)** which is linearly proportional, by proportionality coefficient **a**, to the local corresponding pixel gray value **c(x,y)** in the actin image (Eq. 2):

$$K_{(x,y)} = a * c_{(x,y)} \qquad (2)$$

Microtubules and other organelles are known to bear intra-cellular compression forces. Since we lacked experimental data to represent these based on direct experimental observation, we decided to generate a mechanically coherent intracellular compression network between the nodes of the model by individually surrounding them with small rigid impenetrable spherical contactors 0.58 µm in diameter. When brought the spherical contactors in contact with one another, they interacted by generating repulsive forces. This method thus generated an intracellular network of compression forces opposed to the actin tension network. To achieve this compression bearing function, we coupled sphere contactors to a frictionless contact law as follows, where **g** is the gap between the contactors and **$R_N$** is the normal reaction force respected all following rules simultaneously.

$$\begin{aligned} &For\ g\ \geq\ 0\ \&\ R_N\ \geq\ 0 \\ &If\ g = 0, R_N \geq 0 \qquad (3) \\ &If\ g > 0, R_N = 0 \end{aligned}$$

All these tensile and compressive interactions between nodes formed a pre-stressed network representing the contractile cytoskeleton. Computations were led using the open source LMGC90 solver, dedicated to divided medium mechanics and multi interaction systems (Dubois et al. 2006). During computation, all actin nodes were free to move until the whole tensile network reached a mechanical equilibrium, which consequently led to the slight adjustment of the magnitude of local tensile forces. Nonetheless, nodal displacements remained very small and could be neglected. Considering this, the expression of the tension due to actomyosin can be simplified to:

$$g \approx g_0 \Rightarrow T_{(x,y)} \approx -K_{(x,y)} \cdot \varepsilon_0 \approx -a * c_{(x,y)} \cdot \varepsilon_0 \quad (4)$$

To find all $T(x,y)$ values we conducted a parametric analysis by changing the value of **a** and consequently the values of internal tension (Fig. 1). In other words, **a** is the unknow common general factor between the amount of actomyosin (which is assumed to be proportional to the generated planar tension force) and local brightness value $c(x,y)$ of the fluorescent actin network. For each value of **a**, the tensile interactions were created in the model and the overall mechanical state was computed using the LMGC90 code. The model was computed for 600 time-steps at 300 μs for each step. The net pulling force generated by the cell model on its focal adhesions was compared with the pulling force exerted by the observed cell on the microposts. Increasing **a** increased the net pulling force of the model. An iterative process governed by a gradient based solving algorithm was used to find the solution for which, based on the value of **a**, the model pulled on its surrounding with the same overall traction intensity (sum of all traction force magnitudes). In addition, linear interpolation enabled the model's gradient-based algorithm to converge faster towards the solution. Once the overall traction intensity generated by the model was deemed equivalent to the traction force generated by the cell (less than 0.1% difference), we considered that the value of coefficient **a** was acquired and that the modeled actin network was a mechanical equivalent to that of the cell. Moreover, by matching the actin image with the distribution of tension it is possible to directly convert pixel gray values into intracellular tensions and inversely (Manifacier 2016).

After setting **a**, the number of interactions and the force value per interaction given by the model still depended on its spatial resolution. The resolution of the model was determined by $d_0$, the minimum distance between two nodes, which delimited the region modeled by each interaction. Indeed, a low-resolution model ($d_0$ is large) will be modeled by a proportionally low number of interactions, thus implying high force magnitudes per interaction, while a higher resolution model ($d_0$ is low) would be represented by a higher number of interactions, leading to low force magnitudes per interaction.

On the other hand, the planar cross linear density of force, which is the amount of force per unit length perpendicular to the direction of the considered tensile forces, remains the same in both cases, because it is resolution-independent.

Thus, to estimate the intra-cellular density of tension within the model, while remaining independent to the resolution of the model, i.e. the mesh size, we considered $T_{cl(x,y)}$ the cross-linear tension (or cross-linear density of tension) as defined as:

$$T_{cl\,(x,y)} = \frac{T_{(x,y)}}{d_0} \approx -a * c_{(x,y)} \cdot \frac{\varepsilon_0}{d_0} \quad (5)$$

$T_{cl(x,y)}$ represented on Fig. 1, expressed in nN/µm, can also be referred to as in-plane tension (Diz-Muñoz et al., 2013). Consequently, we can define $T_{cl\text{-}max}$ as the cross linear density of force corresponding to the highest value of actin density which relates to the maximal gray value ($c_{max}$ = 1).

$$T_{cl-max} \approx -a * c_{max} \cdot \frac{\varepsilon_0}{d_0} = -a \cdot \frac{\varepsilon_0}{d_0} \quad (6)$$

With $T_{mean}$ being the mean tension value of all actin interactions, we defined $T_{cl\text{-}mean}$ as the cross linear mean inter-tension:

$$T_{cl-mean} = \frac{T_{mean}}{d_0} \quad (7)$$

Then, directly from the actin image of the cell, it is possible to estimate the amount of intracellular tension $T_{\perp D}$ crossing perpendicularly to every segment drawn on an actin image as:

$$T_{\perp D} = T_{cl-max} \cdot \bar{c}_D \cdot D \quad (8)$$

with $D$ the length of the segment and $\bar{c}_D$, the mean gray value along the segment (Manifacier 2016). Eq. 8 allows us to estimate the tension transiting through each stress fiber (or projected cell section), once the mean gray value along its diameter is obtained.

To estimate the total intracellular tension for every cell, we combined computation and imaging results and introduced the index $T_{intracell}$, as:

$$T_{intracell} = T_{cl-max} \cdot \bar{c} \cdot \sqrt{A} \quad (9)$$

with $A$ as the cell area, $\sqrt{A}$ as the edge of the square equivalent to the cell (since the majority of the analyzed cells possessed a square shape rather than a circular one), and $\bar{c}$ as the mean gray value over the whole cell on the actin image.

## 5. Results

### 5.1. Computations of the mechanical state of cultured cells

Most cells have an area of about 500 μm² independently of substrate stiffness (no correlation) and ranged between 113 and 2594 μm² (Fig. 4e). Traction force and substrate stiffness were positively correlated (Fig. 4a).

A quick visual inspection (Fig. 2) indicates that the cell model has an equivalent shape to that of the analyzed cell and that tensile interactions form between all the nodes. It should be noted that traction forces calculated by the model are the same as the one assessed by the experiment in terms of the total sum of amplitudes. Furthermore, we can also observe that the traction forces at the periphery of the model are oriented toward the center of the cell as experimentally observed. In addition, these peripheral forces appear to be more significant than the ones at the center, as was experimentally observed. Since, the previously cited qualitative and quantitative indicators agree with our experimental observations we may assume that the model behaves in a qualitatively coherent manner. On the other hand, extreme adhesion force magnitudes are attenuated in the model.

Fig. 3a shows another cell with traction force field on a very hard substrate. Once again, the model calculated the intracellular force distributions by admitting a linear relationship between actin density and local tension force densities (cross-linear tension values). Using Eq. 9, intracellular tension can be estimated directly from actin image by drawing segments in different parts of the cell (Fig. 3c).

## 5.2. How does micropost bending stiffness affect tension within cells?

Cells were classified in groups by the rigidity of the substrate they adhered to. The averaged traction force of each group shows a positive correlation with substrate stiffness ($R^2=0.90$) (Fig 4a). The values of $T_{cl-mean}$, $T_{cl-max}$ and $T_{intracell}$ averaged in each group, follow the same tendency (Fig. 4). While $T_{intracell}$ is a tension assessment through the whole cell, $T_{cl-mean}$ and $T_{cl-max}$ are assessments of local distribution of tension. Furthermore, the gap between $T_{cl-mean}$ and $T_{cl-max}$ may indicate high heterogeneity in intracellular tension while similar values indicate homogeneity. Comparing the averaged values of $T_{cl-mean}$, $T_{cl-max}$ and $T_{intracell}$ in each group, $T_{cl-max}$ value is around twice $T_{cl-mean}$ and $T_{intracell}$ is around 10 times greater than $T_{cl-max}$ (Fig. 4).

## 5.3. How are traction forces, cell area and intracellular tension related?

Computed intracellular tension indexes ($T_{cl-mean}$, $T_{cl-max}$ and $T_{intracell}$) are positively correlated with the overall traction force of a cell (sum of all force magnitudes exerted by the cell on microposts). Thus, as traction forces on the microposts increases, so does intracellular tension and vice versa. $T_{cl-mean}$, $T_{cl-max}$ and $T_{intracell}$ respectively have a linear correlation coefficient $R^2$ of 0.68, 0.59 and 0.88 in relation to total adhesion forces (Fig. 5). Similarly, when considering the averaged $T_{cl-mean}$, $T_{cl-max}$ and $T_{intracell}$, values for each substrate stiffness condition, we found respective correlation coefficients $R^2$ of 0.97, 0.94 and 0.87 in relation to the total traction force per cell (averaged for each stiffness condition). Interestingly, the calculated constant of the linear regression of $T_{intracell}$ versus total traction is non-null (Fig. 5). This finding suggests that if the cells were not adherent, there would still be residual intracellular tension, i.e. prestress, within the cellular structure. Further analysis of the results indicates that this residual intracellular tension involves cross linear residual tension in all cells of about 1 to 3 nN/μm.

Within each of the substrate rigidity-based cells groups, the largest cell has a spread area that is at least 4 to 5 greater than its smallest counterpart, which is a significant variation in spread area. Our analysis does not indicate that larger cells pull more or less on the microposts, since traction forces do not correlate with cell area ($R^2=0.16$). Our computations gave similar results and lead to the following conclusion: larger cells do not possess higher intracellular tension. Neither intracellular tension, mean nor max cross linear tensions correlate with cell area ($R^2=$ 0.22, $1.10^{-5}$, 0.03 respectively). We find that cell size does not seem to impact cell contractility.

# 6. Discussion

## 6.1. How relevant is cross-linear tension as a mechanical indicator to study tensile loads within the cell's actin network?

We may thus be tempted to use tensile stress as an indicator, yet it is impractical for the following reasons. First, calculating mechanical stress requires knowing the cross-section area through which the pulling force transits, which would require 3D image acquisition and vertical discrimination between the cytoplasm and the actin network.

Secondly, the actin network is made of many intertwined filaments composing an anisotropic material. Unless it is a well-defined stress fiber, conventional microscopy imaging can only express relative filament density in terms of gray values, the direction of filamentary substructures cannot be identified clearly. We may therefore conclude that pixel gray values are relative indicators of the net amount of tensile force transiting through a corresponding cell volume, which is proportional to the number of actomyosin filaments. For these reasons, we therefore consider that the cross-linear tension values calculated from a collection of gray values are relevant.

## 6.2. Relationships between net cell traction force, intracellular tension, cell area and substrate stiffness

Experimental findings have observed that larger cells pulled more on their surroundings (Fu 2010), yet, failed to mention if higher internal tensile stresses were the cause. Here, our model shows that there is no apparent link between intracellular tension and cell area. Others micropost experiments from the authors seem to indicate that large cells appear to exert less force on each micropost (Han 2012). We would nonetheless like to state that such results may be misleading, because the deflection expressed by a micropost is determined by the net force applied on that given post, whereas forces applied on central posts may be balancing each other out

Interestingly, a linear interpolation of the results shown in Fig. 5 suggests the existence of internal tension within the cellular structure despite adhesion, which could be thought of as some kind of "baseline" or "default" tension of about 1 nN/μm. Our current explanation is that this tension may be due to actin cortex contractility, because the latter is present whether if the cell is adherent or not. This "default" cortical tension explains why cells round when in suspension. This "baseline" tension would generate a hydrostatic pressure of 133 Pa (details in supplementary data). This value is in accordance with internal hydrostatic pressure values found in the literature ranging between 40 to 400 Pa (Fischer-Friedrich et al., 2014). Although we only assume that "baseline" contractility is due to the actin cortex, we see that this idea is both quantitatively and qualitatively coherent.

$T_{intracell}$ values are consistent with results published in Milan et al. 2016, respectively averaging 24 and 40 nN for the 11 and 31nN/μm substrates. The relationships between total traction and $T_{intracell}$ with force and substrate stiffness are also consistent to those presented in Milan 2016 reveal the same correlations. In Milan et al. 2016, the optimization seeks to minimize all the differences between the amplitudes of the focal forces measured experimentally and those of the model. In our present study, the optimization seeks to minimize the difference between experimental and computed total force tractions which has the advantage of being far faster computationally. If the model of Milan et al. 2016 may be considered as more precise mechanically, our present model delivers consistent results far faster and its plus side is the mapping of tension from an image of the actin in cell.

# 7. Conclusion

In this study we calculated intracellular cross-linear tension values in cells that could spread on micropost substrates of different rigidities. We reassuringly found that traction forces and intracellular tension (internal stress) were positively correlated. Our results indicate that large cells pull more on their surroundings because of their size and not due to significantly higher tensile stress within the actin network. Furthermore, large scale substrate stiffness appears to allow for the emergence of higher tensile stress within cells.

**Fig1**

Process flow diagram of the method

a) Microscope imaging of the actin network & microposts

b) Measuring adhesion forces via micropost deflection

c) Generating a computer model by transforming image pixels into mechanical nodes and connecting these nodes by appropriate mechanical interactions. This is done by linearly adapting the tensile stiffness between mechanical nodes based on the local amount of actin observed on the image. The proportional increase in stiffness is linearly determined by coefficient a. (NB: during the first calculation run the value of a is randomly assigned)

d) Running the model which calculates intracellular and adhesion forces.

e) Comparing the cumulative sums of all measured and calculated adhesion force magnitudes. If they are different the model is rerun (Fig 1 B) with a different a value until calculated and measured adhesion forces are found to be similar.

f) Cross linear tension values are calculated using the now known intracellular tension.

**Fig2**

Visual results

Left) Fluorescence image adhering cell: actin in green, nucleus in blue, focal adhesions (vinculin & top of microposts) in red, adhesion force drawn as white arrows. Peripheral adhesion forces are directed inward. Adhesion force intensities are stronger at the periphery than at the center. Similarly to micropost based adhesion force measurements, the direction central adhesion forces are not necessarily directed inward (left image).

Right) View of the model: adhesions are labeled in red, strong actin in white-gray and weaker actin in blue. The modeled cell maintains its initial shape. Blue to white gray colors indicated the local level of cross linear tension $T_{cl}$.

**Fig3**

Intracellular forces computed using the image-based model.

a) In-vitro cell cultured on a very hard substrate ((~2.5 μm diameter ~7 μm height, bending stiffness 41 nN/μm). Blue vectors represent traction forces calculated by the model. Cross linear tension values $T_{cl}$ are expressed in green.

b) tensile forces $T_i$ in the model; $T_{cl-mean}$ = 1.24 nN/μm and $T_{cl-max}$= 2.92 nN/μm.

c) intracellular tension $T_{LD}$ estimated in different parts of an observed cell using ImageJ software to measure gray values (Schneider 2012).

**Fig4**

a) Total adhesion force (nN) vs substrate stiffness (nN/μm). b) $T_{cl\,max}$, c) $T_{cl\,mean}$ and d) $T_{intracell}$ vs substrate stiffness. e) Cell area vs substrate stiffness.

**Fig5**

Total internal tensions in cells ($T_{intracell}$) computed using the model vs their traction forces measured experimentally.

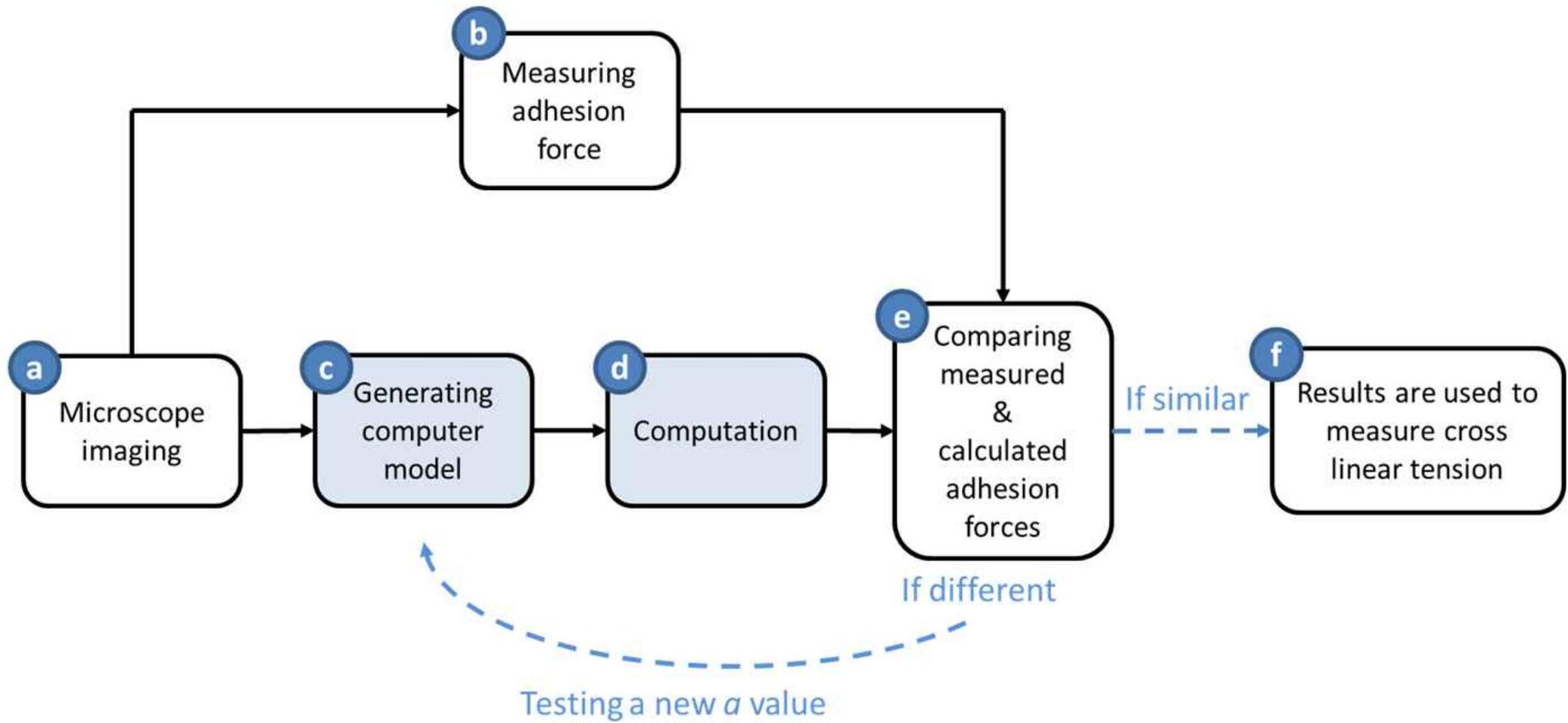

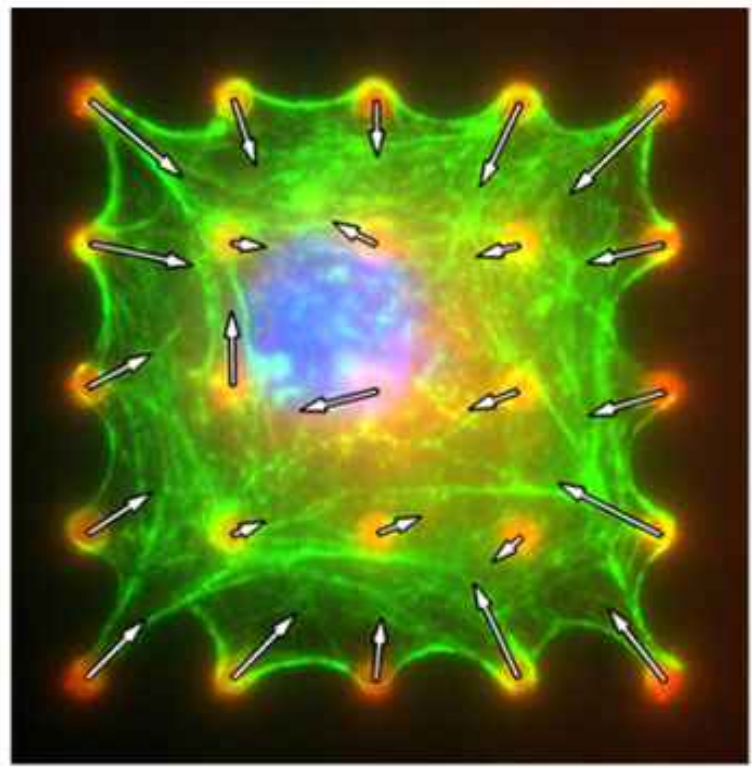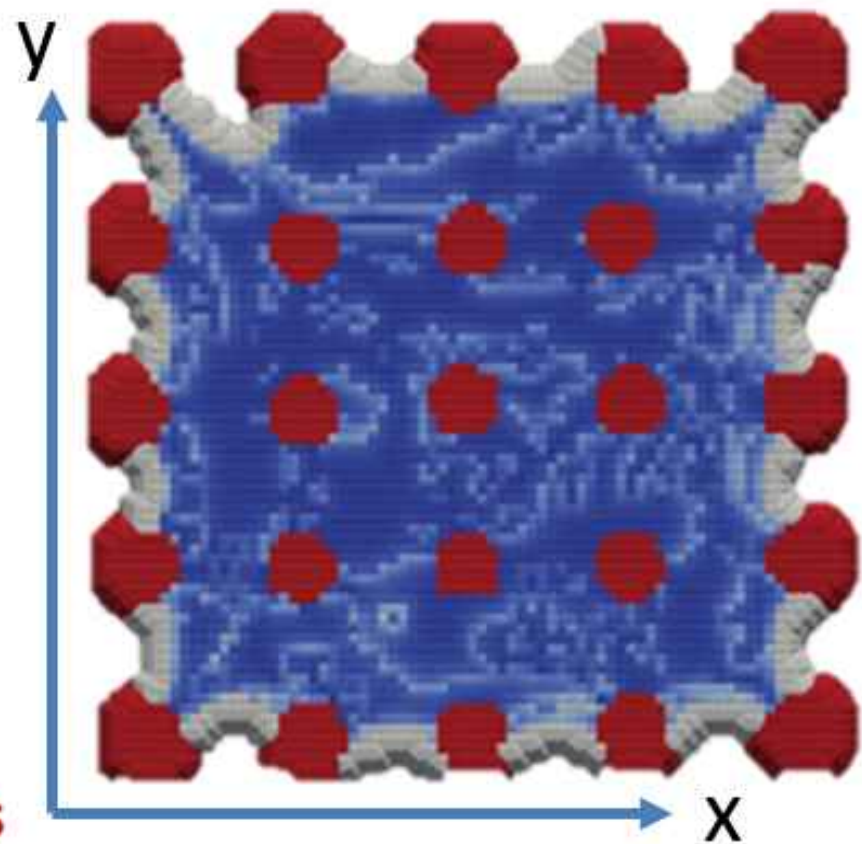

Actin  9 μm  Focal Adhesions

7.2 nN/μm

$T_{cl}(x,y)$

0.7 nN/μm

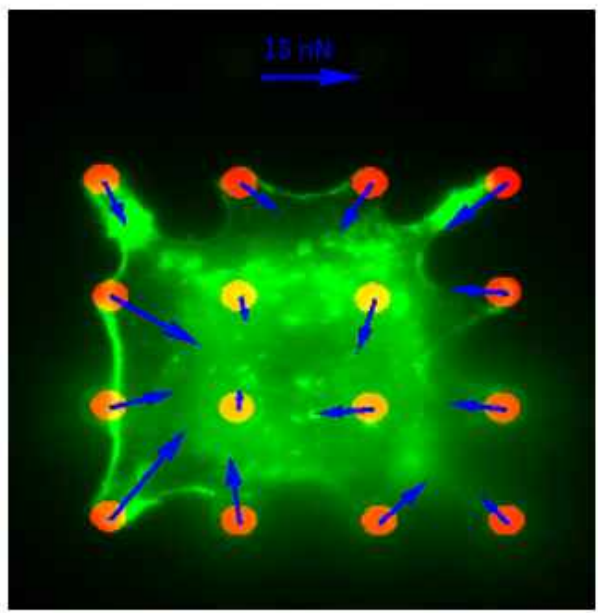
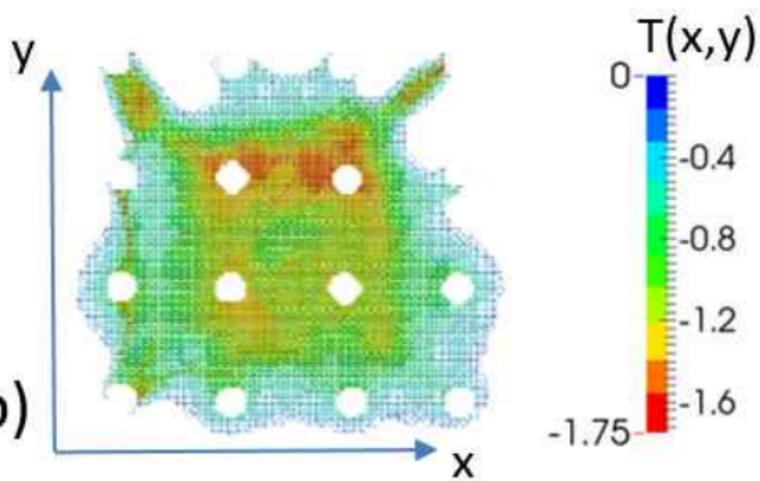
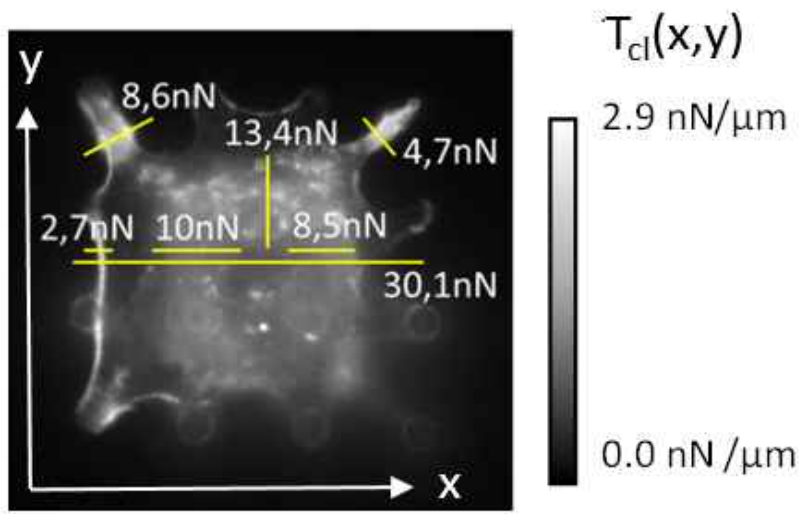

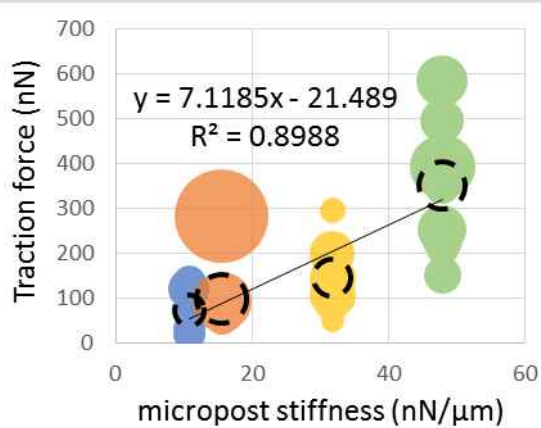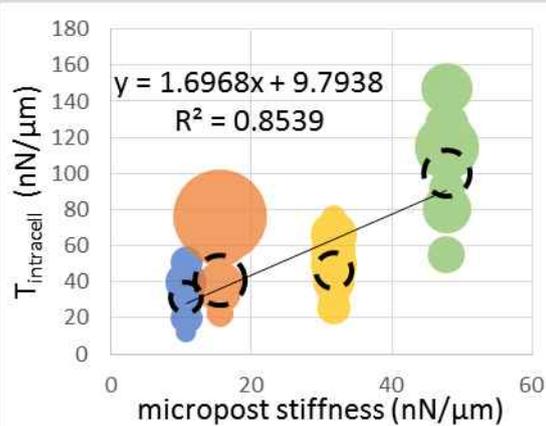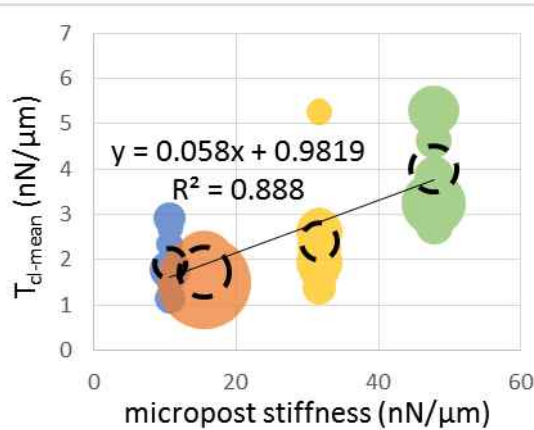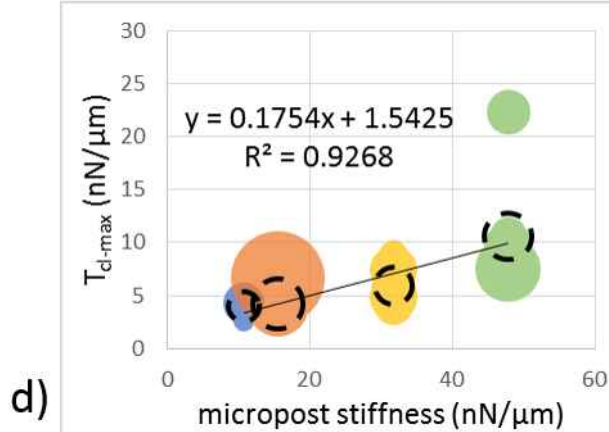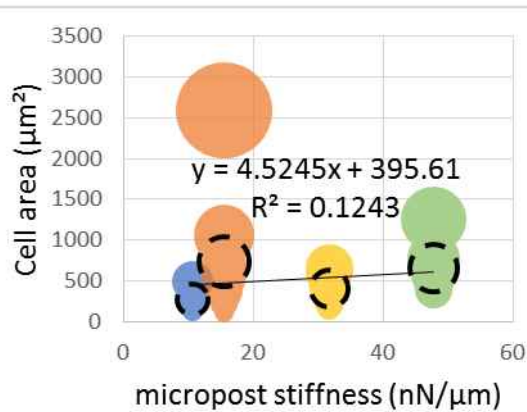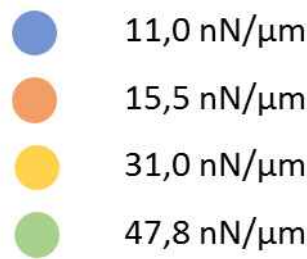

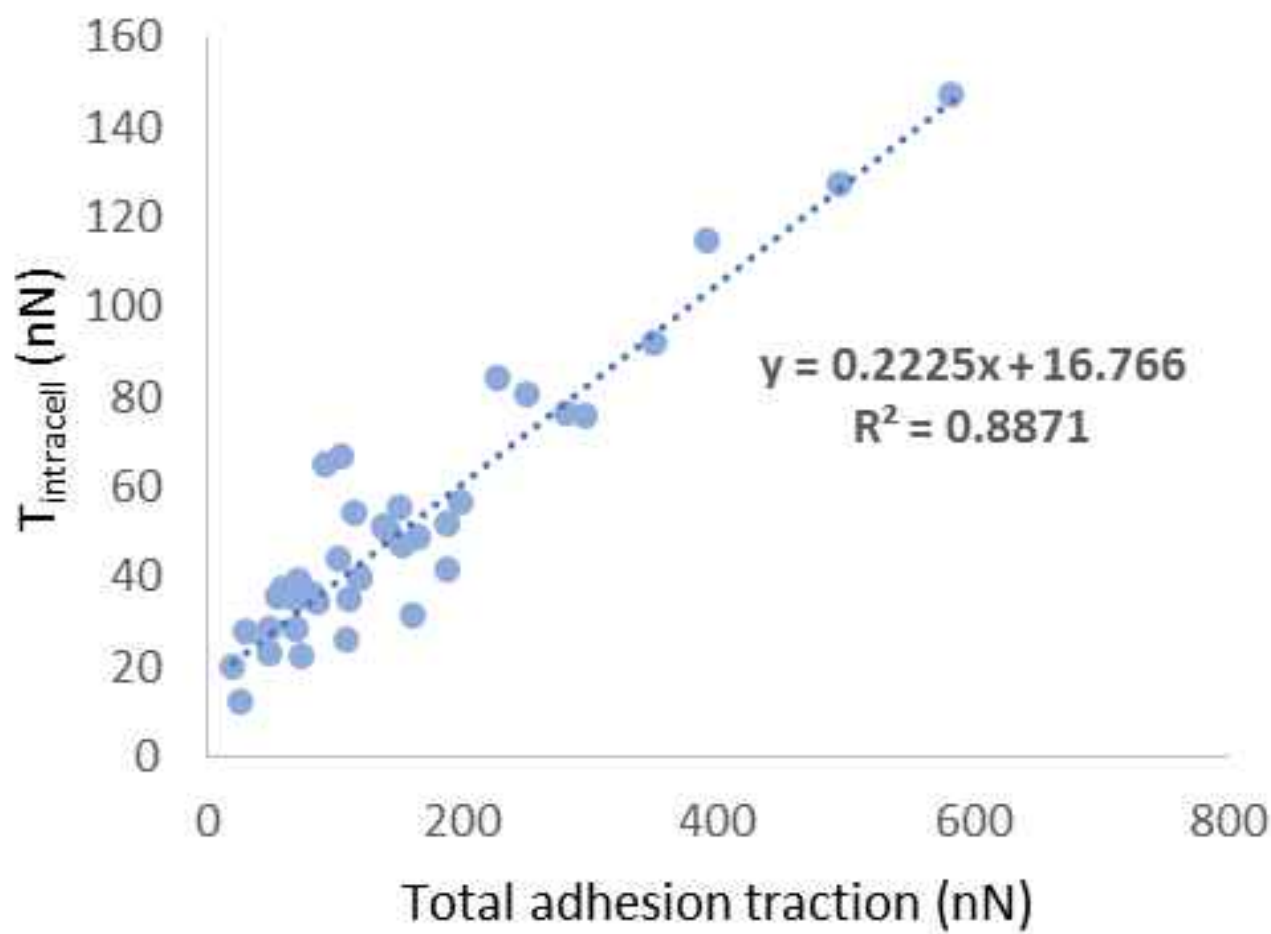

## Supplementary data

### Cell culture and adhesion force measurements

Human pulmonary artery endothelial cells (HPAECs, Lonza) were cultured on four different micropost substrates (Han 2012) manufactured via replica molding of polydimethylsiloxane (PDMS, Sylgard 184, Dow Corning). The microposts were approximated as short beams (short beams (d/L >> 1/10 ). The deflection of a post ($\delta$) was used to determine the local traction force ($F$) of a cell according to Bernoulli's equation $F = k\delta = (3\pi E d^4/64h^3)\delta$ where $k$ is the spring constant of a post, $d$ is its diameter, $h$ is its height, and $E$ is the elastic modulus of PDMS. All four types of substrates were made of the same PDMS with a Young's modulus of 2.5 MPa measured according to ASTM standard D412. Diameters (ranging between 2.14 µm and 2.42 µm) and heights (ranging between 7 µm and 9 µm) resulted in four different bending stiffness (11.0 nN/µm, 15.5 nN/µm, 31.0 nN/µm and 47.8 nN/µm). We categorized these stiffness values into the following designations: Very Soft (11.0 nN/µm ±2.3), Soft (15.5 nN/µm ±3.6), Hard (31.0 nN/µm ±6.2) and Very hard (47.8 nN/µm ±10) (Han 2012). The PDMS microposts presented here have a diameter/height ratio between 0.24 and 0.35> 0.1. Despite the fact that the Bernoulli equation is thus considered inapplicable due to the presence of quadratic forces, previous studies have shown that a linear relationship between micropost deflection and applied force remains a good approximation (Tan2003, Han2012, Fu2010, Lemmon2005, Li2007). Microposts were evenly spaced in a rectangular grid-like fashion, 9 µm from center to center. The edge-to-edge spacing between microposts was therefore ~7 µm. The 7 µm spacing is assumed to be wide enough to prevent ~2 µm long contractile units from being able to perceive micropost bending stiffness (Meacci 2016). Before seeding, in order to confine the spread area of the cells, the tips of micropost substrates were microcontact-printed with fibronectin (50 µg/ml, BD Biosciences) via a stamp-off method (Han 2012; Sniadecki et al. 2013 Methods in Cell Biology) into square shaped pattern areas (441, 900, 1521, or 2304 µm$^2$). Cell seeding density was low enough (<30,000 per mL) so that single cells could separately attach to individual pattern areas. For each of the 4 adhesion conditions, 8-15 cells were imaged and analyzed for traction forces and cytoskeletal tension quantification. The cells were permeabilized using 0.5% Triton for 2 min after being allowed to spread for 14 hours on the micropost arrays. Cells were then stained with Hoechst 33342 (Invitrogen), phalloidin (Invitrogen), IgG anti-vinculin (hVin1, Sigma Aldrich), and anti-IgG antibodies (Invitrogen) with manufacturer-recommended concentration. Images of the cells and microposts were obtained via fluorescence microscopy (Nikon TiE, 60× oil objective, 1.4 NA). We only selected isolated cells. These cells were therefore not able to form cadherin mediated cell-cell junctions with neighboring cells, instead they formed focal adhesions which allowed them to pull on their surrounding environment. Micropost deflection and bending stiffness were used to measure the pulling force exerted by the cell on each post (Lemmon 2005). Micropost deflection was measured by comparing the horizontal position of the center of the top of each micropost to the horizontal position of the base (bottom). The focal plane of the microscope image was manually adjusted to respectively observe the tops and bottoms of the microposts. The difference in position allowed us to calculate the corresponding deflection vector, therefore indicating both the magnitude and directionality of the pulling force exerted on each post. For each observed cell, the traction force was calculated by adding the magnitudes of all adhesion forces. The spreading area of a cell on the micropost array was measured from an outline of its actin image.

**Internal** pressure

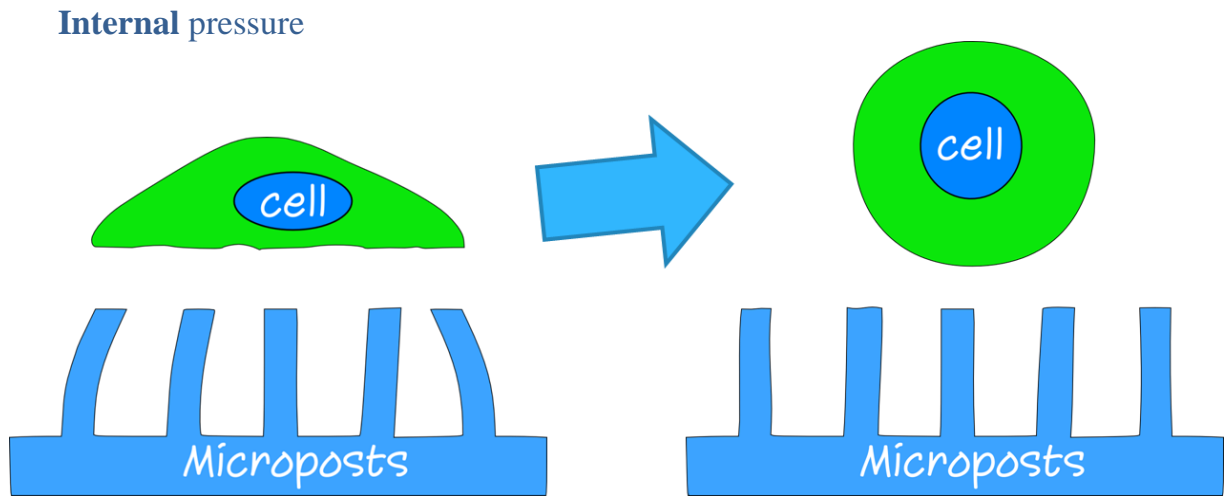

The "default cross linear tension" we find is about 1 nN/μm, however we consider this projected cross linear tension value to be to super impose contribution of the ventral and dorsal actin cortex. We would like to know how this "default" tension could compare to intracellular hydrostatic fluid pressure values expressed in the literature. Since we believe that this default tension is still present in a non-adherent round cell we will conduct our analysis as if the cell was spherical.

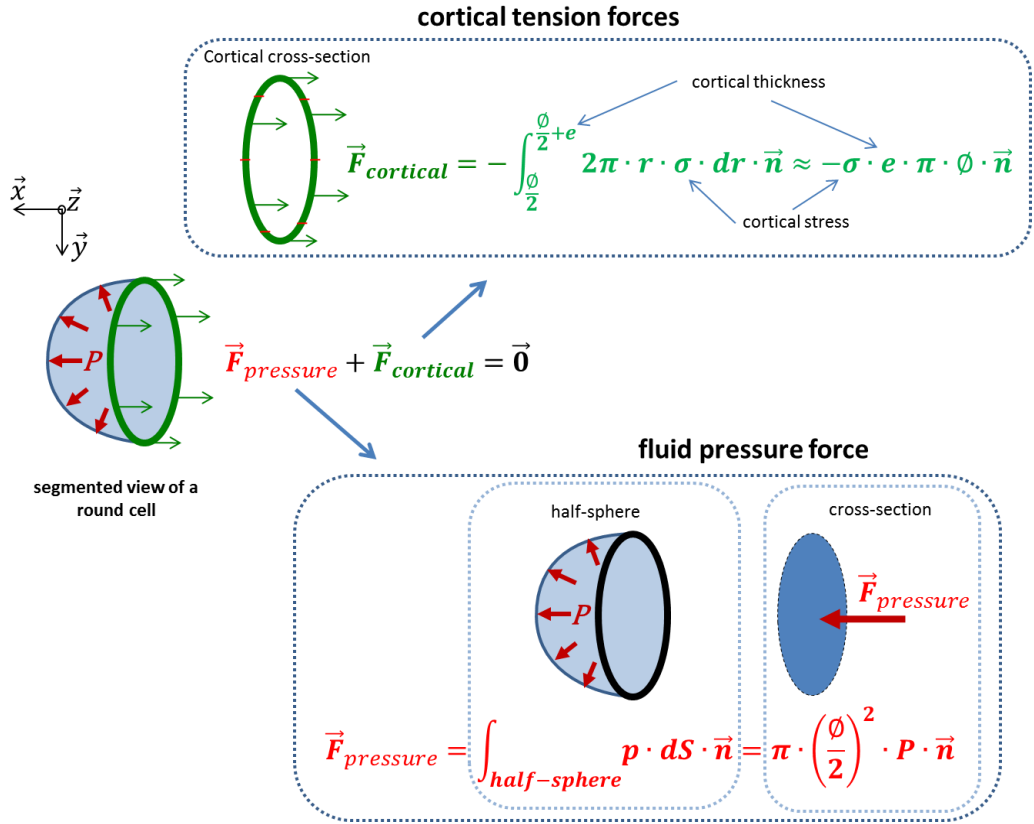

*Figure X: Illustration of cross-sectional forces generated by the actin cortex and the intracytoplasmic fluid pressure of a round cell.*

To calculate the intracellular cytoplasmic pressure inside a round cell, we use the following reasoning. We can consider that if we take a cross section of the cell going straight through its center, we can assume that the peripheral force generated by the thin actin cortex ($\vec{F}_{cortical}$) is counter balanced by the fluid pressure (**P**), and therefore equal and opposite to the fluid pressure force ($\vec{F}_{pressure}$) exerted on the half-sphere (Figure).

*Equation 1*

$$\vec{F}_{pressure} + \vec{F}_{cortical} = \vec{0}$$

In other words, since the actin cortex is 'thin', we can consider that its internal stress (**σ**) is constant. Furthermore, the internal stress multiplied by the thickness (**e**) of the actin cortex is equal to half the default cross linear-tension ($T_{cl-def}$), since this cross-linear tension value accounts for the dorsal and ventral layer of the actin cortex.

*Equation 1*

$$\sigma \cdot e = \frac{T_{cl-def}}{2}$$

*Equation 3*

$$\vec{F}_{cortical} \approx -\sigma \cdot e \cdot \pi \cdot \emptyset \cdot \vec{n} = -\frac{T_{cl-def} \cdot \pi \cdot \emptyset}{2} \cdot \vec{n} \quad \text{with } \emptyset = diameter\ of\ the\ round\ cell$$

*Equation 4*

$$\vec{F}_{pressure} = \pi \cdot \left(\frac{\emptyset}{2}\right)^2 \cdot P \cdot \vec{n} \quad \text{with } \vec{n}: unitery\ vector$$

*Equation 5*

$$\vec{F}_{pressure} + \vec{F}_{cortical} = \vec{0} \quad \rightarrow \quad P = \left|\frac{2 \cdot T_{cl-def}}{\emptyset}\right|$$

Based on our calculations using Equation , for $T_{cl-def}$ equal to ~1 nN/µm and a diameter ($\emptyset$) of 15 µm, we obtain a theoretical[1] hydrostatic pressure ($P$) of 133 Pa. This value is in accordance with internal hydrostatic pressure values found in the literature, which range between 40 to 400 Pa[1]. We can therefore conclude that it would not only be qualitatively coherent, but also quantitatively coherent to assume that the "default" tension found by the image based model corresponds to cortical actin contraction.

1. Fischer-Friedrich, E., Hyman, A. A., Jülicher, F., Müller, D. J. & Helenius, J. Quantification of surface tension and internal pressure generated by single mitotic cells. *Sci. Rep.* **4**, 6213 (2014).

---

[1] The expressed pressure value equates to relative pressure, ambient pressure must be added to obtain absolute pressure.